\documentstyle[sprocl]{article}

\def\NPB{{\em Nucl. Phys.} B}
\def\PLB{{\em Phys. Lett.}  B}
\def\PRL{\em Phys. Rev. Lett.}
\def\PRD{{\em Phys. Rev.} D}

\def\be{\begin{equation}}
\def\ee{\end{equation}}
\def\bea{\begin{eqnarray}}
\def\eea{\end{eqnarray}}

\def\dc{{d^c}}
\def\uc{{u^c}}
\def\ec{{e^c}}

\def\Tr{{\rm Tr}\, }
\def\tr{{\rm tr}\, }

\def\half{\frac{1}{2}}

\def\be{\begin{equation}}
\def\ee{\end{equation}}
\def\beq{\begin{equation} }
\def\eeq{\end{equation} }
\def\beqn{\begin{eqnarray}}
\def\eeqn{\end{eqnarray}}

\def\vev#1{\langle #1\rangle}
\def\mvev#1{|\langle #1\rangle|^2}

\def\etal{{\it et.\ al\/}}

\begin{document}

\rightline{ACT-5/00}
\rightline{CTP-TAMU-09/00}
\rightline{hep-ph/0003208}
\rightline{March 2000}
\rightline{\phantom{March 2000}}

\title{Phenomenological Survey of a\\
 Minimal Superstring Standard Model
\footnote{Talk presented at PASCOS 99, Lake Tahoe, CA December
10--16 1999.}
}

\author{Gerald B. Cleaver}

\address{Center for Theoretical Physics,
Department of Physics\\
Texas A\&M University,
College Station, Texas 77843\\
and\\
Astro Particle Physics Group\\
Houston Advanced Research Center (HARC)\\
The Woodlands, Texas 77381}

\def\eps{\epsilon }
\def\Tr{{\rm Tr}\, }
\def\tr{{\rm tr}\, }
\def\vev#1{\langle #1\rangle}
\def\mvev#1{|\langle #1\rangle|^2}
\def\Q#1{Q_{#1}}
\def\dc#1{d^{c}_{#1}}
\def\uc#1{u^{c}_{#1}}
\def\h#1{h_{#1}}
\def\hb#1{{\bar{h}}_{#1}}
\def\L#1{L_{#1}}
\def\ec#1{e^{c}_{#1}}
\def\Nc#1{N^{c}_{#1}}
\def\H#1{H_{#1}}
\def\V#1{V_{#1}}
\def\Hs#1{H^{s}_{#1}}
\def\Vs#1{V^{s}_{#1}}
\def\p#1{\Phi_{#1}}
\def\pp#1{\Phi^{'}_{#1}}
\def\pb#1{{\overline{\Phi}}_{#1}}
\def\pbp#1{{\overline{\Phi}}^{'}_{#1}}
\def\ppb#1{{\overline{\Phi}}^{'}_{#1}}

\maketitle\abstracts{We discuss a heterotic--string solution
in which the observable sector effective field theory just below 
the string scale reduces to that of the MSSM, 
with the standard observable gauge group being just
$SU(3)_C\times SU(2)_L\times U(1)_Y$ and the
$SU(3)_C\times SU(2)_L\times U(1)_Y$--charged 
spectrum of the observable sector consisting solely of the MSSM spectrum.
Associated with this model is a set of distinct flat directions of 
vacuum expectation values (VEVs) of fields
that all produce solely the MSSM spectrum.
Some of these directions only involve VEVs of non--Abelian singlet fields
while others also contain VEVs of non--Abelian charged fields.}

\section{Flat Directions in Three--Generation Heterotic--String Models}

Over the past decade the free fermionic formulation\cite{fff} of the
heterotic string has been utilized to derive the most realistic\cite{real,fny}
string models to date. In some of these models the observable 
sector gauge group directly below the string scale is a Grand Unified Theory 
while in others it is the (MS)SM gauge group, 
$SU(3)_C\times SU(2)_L\times U(1)_Y$, joined by a few extra $U(1)$ symmetries.
In chiral three generation models of the latter class, one of the additional 
Abelian gauge groups is inevitably anomalous. That is, the trace of the
$U(1)_A$ charge, $\Tr Q^{(A)}$, generates a Fayet--Iliopoulos (FI) term, 
\beqn
\eps\equiv\frac{g^2_s M_P^2}{192\pi^2}\Tr Q^{(A)}.
\label{fit}
\eeqn
The FI term
breaks supersymmetry near the Planck scale, and destabilizes
the string vacuum. Supersymmetry is restored and the vacuum is
stabilized if there exists a direction, ${\phi}=\sum_i\alpha_i \phi_i$,
in the scalar potential which is $F$--flat and also
$D$--flat with respect to the non--anomalous gauge symmetries
and in which $\sum_i Q_i^A\vert\alpha_i\vert^2$ and $\eps$
are of opposite sign.
If such a direction exists it will acquire a vacuum expectation
value (VEV)
cancelling the anomalous $D$--term, restoring supersymmetry and
stabilizing the string vacuum. Since the fields
corresponding to such a flat direction typically
also carry charges for the non--anomalous $D$--terms, a
non--trivial set of constraints on the possible choices of VEVs is imposed:
\beqn
\vev{D_{A}} &=& \sum_m Q^{(A)}_m |\vev{\varphi_{m}}|^2 +
\eps  = 0\,\, , \label{daf}\\
\vev{D_i} &=& \sum_m Q^{(i)}_m |\vev{\varphi_{m}}|^2 = 0\,\, ,
\label{dana}\\
\vev{D_a^{\alpha}}&=& \sum_m 
\vev{\varphi_{m}^{\dagger} T^{\alpha}_a \varphi_m} = 0\,\, , 
\label{dtgen} 
\eeqn
with $T^{\alpha}_a$ a matrix generator of the non--Abelian 
gauge group $g_{\alpha}$ 
for the representation $\varphi_m$. 
These scalar
VEVs will in general break some, or all, of the additional
symmetries spontaneously. 
 
Additionally one must insure that
the supersymmetric vacuum is also $F$--flat. 
Each superfield $\Phi_{m}$ (containing a scalar field $\varphi_{m}$
and chiral spin--$\half$ superpartner $\psi_m$) that appears
in the superpotential imposes further constraints on the scalar VEVs. 
$F$--flatness will be broken (thereby destroying spacetime supersymmetry) at 
the scale of the VEVs unless,
\beq
\vev{F_{m}} \equiv \vev{\frac{\partial W}
{\partial \Phi_{m}}} = 0; \,\, \vev{W}  =0,
\label{ff}
\eeq
where $W$ is the superpotential which contains cubic level and
higher order non--renormalizable terms. 
The higher order terms have the generic
form $$<\Phi_1^f\Phi_2^f\Phi_3^b\cdots \Phi_N^b>.$$ Some of the
fields appearing in the non--renormalizable terms will
in general acquire a non--vanishing VEV by the anomalous
$U(1)$ cancellation mechanism. Thus, in this process some of the 
non--renormalizable terms induce effective renormalizable
operators in the effective low energy field theory 
wherein either all fields or all fields but one are replaced with VEVs.  
One must insure that such terms do not violate supersymmetry
at an unacceptable level.  

\section{A Minimal Superstring Standard Model}

In addition to possessing an anomalous $U(1)_A$ symmetry,
chiral three generation   
$SU(3)_C\times SU(2)_L\times U(1)_Y\times \prod_i U(1)_i$ 
models generically contain numerous non--MSSM  
$SU(3)_C\times SU(2)_L\times U(1)_Y$--charged states,
some of which only carry  
MSSM$\times \prod_i U(1)_i$ 
charges and others of which also carry hidden sector (non)--Abelian
charges. Recently, we showed that in some of these models 
it is actually possible to decouple all such non--MSSM states from the low
energy effective field theory. 
For example, in the ``FNY'' model first presented in Ref.\ 3,
we discovered there are
several flat directions\cite{cfn1,cfn2} for which almost all
MSSM--charged exotics receive FI masses
(typically of the scale $5\times 10^{16}$ GeV to $1\times 10^{17}$ GeV)
while the remaining MSSM--charged exotics (usually composed of simply a
$SU(3)_C$ triplet/anti--triplet pair) acquire masses slightly suppressed 
below the FI scale by a factor of ${\cal O}(\frac{1}{10}$ to $\frac{1}{100})$.
Some of our flat directions accomplishing this feat contain only 
VEVs of non--Abelian singlet fields\cite{cfn2,cfn3}
while others of ours also contain VEVs of non--Abelian charged 
fields\cite{cfn4,cfn5}. Along these directions,
exactly three generations of 
($\Q{i}$, $\uc{i}$, $\dc{i}$, $\L{i}$, $\ec{i}$, $\Nc{i}$)
fields and a pair of electroweak Higgs, $h_u$ and $h_d$, are the only
MSSM--charged fields that remain massless significantly below the FI scale. 
The non--MSSM--charged singlet fields 
and hidden sector non--Abelian fields that also 
remain massless below the FI scale vary with the flat direction chosen. 

The complete massless spectrums produced, respectively,
by four singlet flat directions were presented in Ref.\ 6.
Detailed analysis of the associated three generation mass textures 
was also performed therein. 
The leading mass terms were found to be
$Q_1u_1^c{\bar h}_1$ and $Q_3d_3^ch_3$ for all four directions. 
The non--Abelian directions considered in Ref.\ 7
produced similar results.
This presented a phenomenological difficulty for 
our first few MSSM--producing flat directions by
implying that the left--handed component of the top and bottom quarks 
live in different multiplets. 
A more detailed study of the mass textures possible from
non--Abelian flat directions is underway and will be presented in Ref.\ 9.

\section{Phenomenogolical Implications}

A string--derived three generation MSSM low energy effective field theory 
resulting from the decoupling of all MSSM--charged exotics via an 
anomaly--cancelling flat direction possesses some unique phenomenological 
characteristics (independent of the viability of the associated
MSSM mass matrices).
For instance, we found that at least one $U_i$ combination usually remains
unbroken by a given flat direction. Generally the surviving extra Abelian 
symmetries could not have been embedded in $SO(10)$ or $E_6$ 
GUTS\cite{cfn5}. 
Relatedly, a common feature in the surviving $U_i$ combinations 
is a flavor non--universality.
Thus, the distinctive collider signatures
of a $Z^\prime$ arising from one such symmetry
will be a non--universality in the production of the different
generations. An additional $Z^\prime$ of this type 
has also been suggested as playing a role
in suppressing proton decay in supersymmetric extensions
of the Standard Model\cite{pati}.

\section*{Acknowledgments}
G.C. thanks the organizers of PASCOS '99 for an excellent conference. 
The work discussed herein was done in collaboration
with A.E.\ Faraggi, D.V.\ Nanopoulos, \& J.W.\ Walker.

\section*{References}
\def\NPB#1#2#3{{\it Nucl.\ Phys.}\/ {\bf B#1} (#2) #3}
\def\PLB#1#2#3{{\it Phys.\ Lett.}\/ {\bf B#1} (#2) #3}
\def\PRD#1#2#3{{\it Phys.\ Rev.}\/ {\bf D#1} (#2) #3}
\def\PRL#1#2#3{{\it Phys.\ Rev.\ Lett.}\/ {\bf #1} (#2) #3}
\def\PRT#1#2#3{{\it Phys.\ Rep.}\/ {\bf#1} (#2) #3}
\def\MODA#1#2#3{{\it Mod.\ Phys.\ Lett.}\/ {\bf A#1} (#2) #3}
\def\IJMP#1#2#3{{\it Int.\ J.\ Mod.\ Phys.}\/ {\bf A#1} (#2) #3}
\def\nuvc#1#2#3{{\it Nuovo Cimento}\/ {\bf #1A} (#2) #3}
\def\RPP#1#2#3{{\it Rept.\ Prog.\ Phys.}\/ {\bf #1} (#2) #3}
\def\etal{{\it et.\ al\/}}

\def\bg{\bibitem}

\end{document}